\documentclass[showpacs,preprintnumbers,amssymb,twocolumn]{revtex4-2}

\usepackage{graphicx}
\usepackage{dcolumn}
\usepackage[dvipsnames]{xcolor}
\usepackage{bm}
\usepackage{amsmath}
\usepackage{amssymb}
\usepackage{epsfig}
\usepackage{amsfonts}
\usepackage{lineno,hyperref}
\usepackage{array}
\usepackage{float}
\usepackage{microtype}
\usepackage{multirow}
\usepackage{adjustbox}
\usepackage[english]{babel}
\usepackage{epstopdf}
\usepackage{blindtext}
\usepackage{subcaption}
\usepackage[a4paper, total={6.5in, 10in}]{geometry}
\usepackage{appendix}

\def \a{\alpha}

\def \o{\omega}

\def \be{\begin{equation}}
\def \ee{\end{equation}}
\def \ben{\begin{eqnarray}}
\def \een{\end{eqnarray}}

\def \p{\bar{p}}

\begin{document}
\title{Particle production rate for a dynamical system using the path integral approach }

\author{Samit Ganguly}
\email{samitganguly1994@gmail.com}
\affiliation{Department of Physics, Haldia Government College, Haldia, Purba Medinipur 721657, India $\&$ \\ Department of Physics, University of Calcutta, 92, A.P.C. Road, Kolkata-700009, India}

\author{Narayan Banerjee}
\email{narayan@iiserkol.ac.in}
\affiliation{Department of Physical Sciences, Indian Institute of Science Education and Research Kolkata, Mohanpur, Nadia, West Bengal 741246, India}

\author{Abhijit Bhattacharyya}
\email{abhattacharyyacu@gmail.com}
\affiliation{Department of Physics, University of Calcutta, 92, A.P.C. Road, Kolkata-700009, India}

\author{Goutam Manna$^a$}
\email{goutammanna.pkc@gmail.com \\$^a$Corresponding author}
\affiliation{Department of Physics, Prabhat Kumar College, Contai, Purba Medinipur 721404, India $\&$\\ Institute of Astronomy Space and Earth Science, Kolkata 700054, India}

\date{\today}

\begin{abstract}
In this work, we investigate the particle creation rate in a dynamical (Vaidya) spacetime using Feynman's path integral formalism within the framework of the effective action approach. We examine three distinct cases involving the following mass functions, each representing dynamical geometries: (i) $m(v,r)=\mu v$, (ii) $m(v,r)=\mu v +\nu r$, and (iii) $m(v,r)=\mu v -\frac{\mu^2 v^2}{2r}$, where $\mu$ and $\nu$ are positive constants that satisfy all known energy conditions. We analyze particle production rates in the region of dynamical horizons, revealing an initial high rate followed by a rapid decline in all cases. Additionally, we explore the thermodynamic properties by calculating the surface gravity and corresponding Hayward-Kodama temperatures for each scenario. Graphical representations show the variation of surface gravity over time for the three cases, offering insights into the system's thermodynamic evolution. Our research investigates the connection between background geometry and the particle creation process, placing it within the broader context of quantum field theory in curved spacetime. The non-stationary nature of Vaidya geometry is highlighted as a valuable framework for examining the dynamic aspects of particle creation. This in-depth analysis enhances our understanding of quantum processes in curved spacetime and may offer insights relevant to thermodynamics and studies of gravitational collapse.

\end{abstract}

\maketitle

\section{Introduction}
Particle production takes place inside the spacetime of black holes, as predicted by Hawking . Hawking postulated that the strong gravitational fields near the event horizon of a black hole had the potential to create particle-antiparticle 
pairs originating from the vacuum \cite{Hawking1,Hawking2}. Usually, these pairs would rapidly annihilate each other. Nevertheless, near the event horizon, one particle has the possibility of falling into the black hole while the other 
can escape, thus avoid their immediate absorption into each other. The particle that has been ejected from the black hole manifests as radiation, often referred to as Hawking radiation. The prediction made by Hawking about the production 
of particles in the spacetime of a black hole serves as a source of inspiration for investigations into the quantum nature of gravity. An analysis of the Hawking effect, involving quantum fluctuations, explores the likelihood of positive-energy 
modes tunneling outward or negative-energy modes tunneling inward through the event horizon, as discussed in Ref. \cite{Parikh,Mitra}. Wondrak et al. \cite{Wondark} have recently provided an analysis of the formation of gravitational pairs and the 
evaporation of black holes using a heat-kernel technique that is similar to the Schwinger effect \cite{Euler,Schwinger}. The researchers have used this approach to analyze an uncharged massless scalar field in Schwarzschild spacetime {\cite{Wondark}. Their findings demonstrate that the curvature of spacetime plays a comparable function to the strength of the electric field in the Schwinger effect. This had been interpreted as the generation of local pairs inside 
a gravitational field and have derived a profile of radial production. The emission peaks are located near the unstable photon orbit. Additionally, a comparison has been made between the particle number and energy flow in the Hawking scenario, revealing that both effects exhibit a comparable magnitude. However, the method for producing these pairs does not explicitly rely on the existence of a black hole event horizon.

We know that in the realm of quantum mechanics, a vacuum state is filled with virtual particle pairs that continuously appear and disappear through spontaneous creation and annihilation processes. These quantum fluctuations have the potential to manifest as actual particle pairs when a background field is present. One notable instance of this phenomenon is the Schwinger effect, which predicts the formation of charged particle pairs when an electric field is present \cite{Euler, Schwinger, Dunne1, Dunne2}. In this method, the particles of a spontaneously created virtual pair are accelerated in opposite directions by the external field. The Heisenberg uncertainty principle allows virtual particles to become real if they collect enough energy over a Compton wavelength to follow the relativistic energy-momentum relation $E^{2}=m^{2}+\vec{p}^{2}$ during their separation. Strong electric fields can produce the pair in various contexts, such as high-intensity laser beams \cite{Kohl, Fedotov}, ultraperipheral heavy-ion collisions, exotic atoms \cite{Paul}, etc. A Schwinger-analog effect has been discovered in graphene \cite{Berdyugin, Katsnelson1, Katsnelson2} in the context of condensed matter physics. Other particle creation mechanisms may be found in \cite{Schafer, Ford, Zeldovich1, Zeldovich2, Zeldovich3, Zeldovich4, Parker, Birrell, Verdaguer, Mukhanov, Parker1, Vassilevich} within the realm of cosmology, small time-dependent anisotropies, or self-interaction of the quantum field.

Our objective is to explore the particle production rate in a dynamical spacetime by applying the effective action principle through the path integral method, as outlined in \cite{Vassilevich, Verdaguer}. In the second section, we provide a brief description of Vaidya geometry and the associated dynamical horizons. In section three, we explore the effective action principle and its potential applications in dynamical spacetime. In section IV, we derive the particle production rate using the effective action principle via the path integral method for three distinct generalized Vaidya mass functions and analyze the results. The final section presents our conclusions.

\section{Vaidya Metric} 
In 1951, Vaidya found out the first relativistic line element that describes the spacetime of a radiating star \cite{Vaidya, Visser, Vaidya1, Vaidya2, Vaidya3, Vaidya4, Vaidya5}.  The primary difference between the Schwarzschild solution and the Vaidya solution is that the former describes a static spacetime with a constant mass and the latter leads to a nonstatic scenario with a mass function that varies with time.

Husain \cite{Husain} extended the conventional Vaidya spacetime by for a null fluid. Wang and Wu \cite{Wang} further showed that the energy-momentum tensor components for Type-I and Type-II matter fields are proportional 
to the mass function. The off-diagonal term in the generalized Vaidy metric could lead to negative energy for a particle \cite{Vertogradov}. For a specified mass function, the generalized Vaidya metric produces a homothetic Killing 
vector, providing additional symmetry that could aid in constructing a constant of motion related to both angular momentum and energy \cite{Vertogradov5}. Consequently, the field equations are satisfied by combining specific 
solutions \cite{Cahill}. In this broader framework, the mass function depends on both space ($r$) and time ($v$). The line element of the generalized Vaidya metric can be expressed as
\ben
 ds^2 = -\Big(1 - \frac{2m(v,r)}{r}\Big) dv^2 + 2dvdr + r^2 d\Omega^2
 \label{1}
\een
where $d\Omega^2$  is the metric on the unit $2-$sphere, $m(v,r)$ is the generalized mass function, $r$ is the radial coordinate, $v$ represents the advanced time (for ingoing radiation) or retarded time (for outgoing radiation). If $m=m(v)$, 
and not a function of $r$, the metric (\ref{1}) reduces to the original Vaidya metric. Several works can be found on (generalized) Vaidya spacetime within the domains of general relativity and astrophysics ~\cite{Joshi1,Joshi2,Joshi3,Joshi4, Joshi5,Joshi6,Oppenheimer,Malafarina,gm1,gm2,gm3}.

The horizon of the Vaidya spacetime is dynamic and depends on the mass function which is a function of time and position.

The energy and angular momentum fluxes carried by gravitational waves across the dynamical horizons, as well as the 
equation that describes the variation in the dynamical horizon radius, can be found in the Refs.: \cite{ashtekar1,ashtekar2,ashtekar3,ashtekar4,hayward,badri}. The definition of {\it dynamical horizon} is the following.

A smooth, three-dimensional space-like submanifold $H$ in spacetime $\cal{M}$ is said to be a dynamical horizon if it can be foliated by a family of closed $2-$surfaces such that, on each leaf $S$, the expansion $\Theta_{(l)}$ of 
one null normal $l^{a}$ vanishes and the expansion $\Theta_{(n)}$ of the other null normal $n^{a}$ is strictly negative. And {\it the modified definition proposed by Sawayama \cite{Sawayama}} is

A smooth, three-dimensional, spacelike or \textbf{\textit{timelike}} submanifold $H$ in a space-time is said to be a dynamical horizon if it is foliated by a preferred family of 2-spheres such that, on each leaf $S$, the expansion $\Theta_{(l)}$  of a 
null normal $l^{a}$ vanishes and the expansion $\Theta_{(n)}$ of the other null normal $n^{a}$ is strictly negative. 

For an example \cite{blau} in Minkowski spacetime, 
(i) radially outgoing light rays, $l =\partial_{v},~ v = t + r$, have expansion

$\Theta_{l}=\nabla_{\a}(\partial_{v})^{\a}=\frac{1}{r^2}\p_{\a}\Big(r^{2}(\p_{t}+\p_{r})^{\a}\Big)=+\frac{2}{r}>0$, 
and (ii) radially ingoing light rays $n=\p_{u}, u = t-r$, have expansion

$\Theta_{n}=\nabla_{\a}(\partial_{u})^{\a}=\frac{1}{r^2}\p_{\a}\Big(r^{2}(\p_{t}-\p_{r})^{\a}\Big)=-\frac{2}{r}<0$, which indicate that outgoing light rays expand while ingoing light rays contract.
Following \cite{ashtekar3, gm3},  in the concept of world tubes, if the marginally trapped tubes (MTT) is 
\begin{enumerate}
\item [1.] spacelike, then it is called a dynamical horizon (DH), and under some conditions, it provides a quasi-local representation of an evolving black hole.
 
\item [2.] timelike, then the causal curves can transverse it in both inward and outward directions, where it does not represent the surface of a black hole in any useful sense, it is called a timelike membrane (TLM).

\item [3.] null, then it describes a quasi-local description of a black hole in equilibrium and is called an isolated horizon (IH).
\end{enumerate}

\section{Effective Action}

Effective action is a notion in quantum field theory that provides an understanding of the behavior of quantum fields at varying energy scales \cite{Birrell, Mukhanov, Parker}. This statement refers to the incorporation of quantum fluctuations or quantum corrections in the claasical action. The work of Vassilevich \cite{Vassilevich} discusses the application of the heat kernel approach to quantum field theory or quantum gravity to characterize the 
effective action. Fock et al. (1937) \cite{Fock} observed that it is advantageous to express Green functions as integrals over an auxiliary coordinate, known as the ``proper time," of a kernel that obeys the heat diffusion equation. Subsequently, 
Schwinger \cite{Schwinger} acknowledged that this particular representation provides more clarity on certain matters such as renormalization and gauge invariance in the presence of external fields. These two papers brought the concept of the 
heat kernel to the field of quantum theory. DeWitt \cite{DeWitt1, DeWitt2, DeWitt3, DeWitt4} used the heat kernel as the main tool in his approach to quantum field theory and covariant quantum gravity. This approach became very useful over a long period of time.

The effective action, often denoted by $\Gamma[\phi]$, is a functional of the classical fields, which are themselves expectation values of the quantum fields in the presence of sources. It is obtained by integrating out the quantum fluctuations 
around the classical field configuration ($\phi$) in the path integral formalism \cite{blau, Mukhanov, Vassilevich, Fock, Schwinger}. Mathematically, it can be expressed as,
\ben
\frac{\delta\Gamma}{\delta \phi}=-J.
\label{2}
\een

In particular, in the absence of the external source (i.e., for $J(x)\to 0$) we have $\frac{\delta\Gamma}{\delta \phi}=0$. The effective action governing a quantum field in a background spacetime and its variation gives
the equation of motion for the field with quantum corrections. However, it is well-defined only for flat or static spacetimes and for a constant field. But effective action is for even broader circumstances, allowing for dynamical background 
spacetimes and background fields. Its variation produces dynamical equations for both the field and the background spacetime in a self-consistent way \cite{Verdaguer}. A key feature of the effective action is that it generates 
one-particle-irreducible (1PI) Green functions upon functional differentiation. These Green functions are crucial for understanding the interactions between particles in the presence of a background field \cite{BARVINSKY1985}.\\

The in-out effective action formalism is widely used in various areas of theoretical physics, including the study of quantum fields in curved spacetime, quantum cosmology, and non-equilibrium quantum field theory. It helps us understand things like the Swinger effect (how particles are created in strong fields) \cite{Schwinger}, Hawking radiation (how quantum corrections affect black hole evaporation) \cite{Hawking1, Hawking2}, and the dynamics of phase transitions in the early universe \cite{Bastianelli}. Our focus is especially on understanding the dynamics of particle formation in dynamical spacetime. The ``in-out" formalism involves computing the vacuum persistence amplitude, which gives the probability of the number of particles being created from the vacuum. The vacuum persistence amplitude in the presence of an external field is given as: $\Big<0_{out}|0_{in}\Big>=e^{i\Gamma[g_{\mu\nu}]}$
where $\Gamma[g_{\mu\nu}]$ is the effective action evaluated in the presence of the external field. The imaginary part of the effective action, $\textit{Im}(\Gamma)$, is directly related to the probability of particle creation, which is mathematically expressed as: $P_{pair}\propto exp(-2 Im (\Gamma))$. An imaginary non-zero part indicates that particle creation is occurring.

One common method to approximate the effective action in a non-static background is through the use of perturbation theory and loop expansions, where higher-order loops account for more significant quantum corrections \cite{Gilkey}. This approach, however, can become complex in strongly coupled theories or near critical points where non-perturbative effects are significant. It should be important to note that defining a vacuum state is a considerable problem when building QFT in time-dependent settings \cite{Verdaguer}. There are a few scenarios in which vacuum states in QFT in dynamic spacetimes may be clearly defined: (1) the term statically bounded or asymptotically stationary refers to spacetimes in which it is assumed that at $t=\pm\infty$, the background spacetime becomes stationary and the background fields remain constant; (2) conformally invariant fields in conformally static spacetimes. In these two examples, the Fock spaces are well-defined, allowing for the calculation of particle amplitudes using an S-matrix. (3) It is easy to understand and use a simple method to describe the so-called (nth-order) adiabatic vacuum or number state if the background spacetime does not change too quickly. In our case, we consider the third possibility to specify the limit to use the path integral technique in our investigation.

\section{Particle Production in Vaidya geometry}

In quantum physics, quantum fluctuations allow the creation of particle-antiparticle pairs of virtual particles. These pairs exist for an extremely short time, and then annihilate each other. In some cases, it is possible to separate the pair 
using external energy, therefore they avoid annihilation and become actual (long-lived) particles. For long-lived particle creation (which respects energy conditions) one particle must fall back into the black hole and the other to 
leave to future infinity where an asymptotic observer could detect the particle. Hiscock et al. \cite{Hiscock} conducted a study on the creation of particles by shell-focusing singularities in Vaidya geometry. If the singularity is 
marginally naked, meaning its Cauchy horizon aligns with the event horizon, they have successfully determined the spectrum of particles by Hawking's approach. This dynamic closely resembles those observed in extremal black holes. While marginally naked singularities and extremal black holes display similar characteristics, they fundamentally differ in horizon structure, particle spectra, and stress-energy behavior. In marginally naked singularities, the Cauchy horizon momentarily coincides with the event horizon, exposing infinite curvature, while extremal black holes have a degenerate event horizon that fully cloaks the singularity. Marginally naked singularities produce a quasi-thermal spectrum with frequency-dependent temperature, whereas extremal black holes have zero temperature. Additionally, the stress-energy tensor diverges on the Cauchy horizon of marginally naked singularities but remains finite for extremal black holes. Firouzjaee and Ellis  \cite{Ellis} showed that when considering quantum  fluctuations, particle formation may be predicted by including a generalized uncertainty principle in the context of a slowly varying dynamical horizon. 

Our current objective is to determine the rate at which particles are created using the 
effective action principle through the path integral approach. In this case, we consider the most basic example where a scalar field $(\phi)$ is linked to gravity in a non-minimal way. The action for this field may be expressed 
as \cite{Birrell, Parker, Mukhanov, Verdaguer}
\ben
S[\phi]=\int d^4x \sqrt{-g}\Big[\frac{1}{2}g^{\mu \nu}\partial_{\mu}\phi\partial_{\nu}\phi-\frac{1}{2}m^2\phi^2-\frac{1}{2}\zeta R\phi^2\Big]\nonumber\\
\label{3}
\een
where $\zeta$ is the field-curvature coupling constant. Rewriting the action (\ref{3}) we get,
\ben
S[\phi]=-\int d^4x\sqrt{-g}\frac{1}{2}[\phi(\Box+m^2+\zeta R)\phi]\nonumber \\+ \text{boundary terms}
\label{4}
\een
where $\Box\equiv \nabla^{\mu}\nabla_{\mu}$. Following \cite{Verdaguer}, the effective action for this scalar field connected to the curvature of spacetime may be expressed as a functional of the metric tensor $g_{\mu\nu}$ as
\ben
Z[g_{\mu\nu}]=< out, 0|0, in>=\mathcal{N}\int\mathcal{D}\phi~ e^{iS[\phi,g_{\mu\nu}]}
\label{5}
\een
where $\mathcal{N}$ is a normalization factor that may be expressed in terms of the free component of the action $S_{0}[\phi]$ by setting $\zeta$ equal to zero,
\ben
\mathcal{N}=\frac{1}{\int \mathcal{D}\phi~e^{iS_{0}[\phi]}}.
\label{6}
\een
The generating functional $Z[g_{\mu\nu}]$ may be interpreted as the vacuum amplitude $< out, 0|0, in >$, where the presence of an external source can disrupt the stability of the initial vacuum state, leading to the creation of particles. 
Effective action may be defined as $Z[g_{\mu\nu}]=e^{i\Gamma[g_{\mu\nu}]}$. Consequently, $\Gamma[g_{\mu\nu}]$ will produce all the connected green functions, and it is now defined as
\ben
\Gamma[g_{\mu\nu}]=-iln~(Z[g_{\mu\nu}]).
\label{7}
\een
Neglecting the boundary terms in action (\ref{4}), we can write
\ben
Z[g_{\mu\nu}]&&=e^{i\Gamma[g_{\mu\nu}]}\nonumber\\&&=\int\mathcal{D}\phi~e^{-\frac{i}{2}\int d^4x\sqrt{-g}\frac{1}{2}[\phi(\Box+m^2+\zeta R)\phi]}.
\label{8}
\een
Based on the definition of effective action provided earlier (\ref{7}), it is clear that if the initial and final vacuum states are unstable owing to the existence of the background field, the effective action will become complex. This may be 
expressed as
\ben
|< out, 0|0, in >|^2=e^{-2Im \Gamma}\equiv 1-P
\label{9}
\een
where $P$ represents the probability of the pair production. Thus, the probability of creating a pair of particles from the initial vacuum is proportional to the imaginary part of the effective action assessed at the mean geometry over the history of the universe. When the in-out vacuum expectation value is very small or approximately zero (i.e. $Im \Gamma$ is very large or, $P\approx 1 $), it signifies that the vacuum state has entirely decayed due to particle pair production. The initial vacuum is no longer stable, as particle-antiparticle pairs are continuously created in response to the background field or dynamic spacetime geometry. The instability of the vacuum suggests that backreaction effects become significant. The energy-momentum tensor of the created particles may alter the background spacetime, which in turn may lead to self-consistent dynamics that could modify the rate of particle creation over time indicating a total quantum domination of the physical behavior of the system. In this regime, perturbative approaches break down, placing the phenomenon beyond the scope of standard perturbative analysis. Conversely, when $Im \Gamma=0$, that is, $P=0$, the spacetime curvature or background field configuration fails to provide enough dynamical effects to perturb the initial vacuum state. This leads to a vacuum stability condition where no energy is transferred from the background to the field modes, leading to a complete absence of particle creation processes. However, if the imaginary part is small but finite, we can write
\ben
P \approx 2 Im~\Gamma[g_{\mu\nu}].
\label{10}
\een}
The imaginary component of the one-loop effective action represents the corresponding vacuum non-persistence rate \cite{Vassilevich, BARVINSKY1985}. Once renormalization is performed, it may be defined as follows
\ben
\Gamma= -\frac{\Tilde{\mu}^{2z}}{2}\int d^4x\sqrt{-g}\int_0^{\infty}\frac{ds}{s^{1-z}}\nonumber\\ \times\int_{x(0)=x(s)}\mathcal{D}x(\tau)e^{-S_E[x(\tau)]}.
\label{11}
\een
In the given situation, we use a zeta function regularization technique that involves a mass scale denoted by $\Tilde{\mu}$ and a parameter represented by $z\in\mathbb{C}$ \cite{Wondark, Dunne1}. In the path integral, all excitation paths 
that begin and terminate at the same point are considered. A closed trajectory requires the field excitation to move in both the forward and backward directions in external time. The phenomenon may be seen as the virtual creation and 
annihilation of a pair of particles and their corresponding antiparticles. The coincidence limit in the heat kernel refers to the total of all particle-anti-particle paths that begin and terminate at the same point. The integral may be 
divided into two parts: the classical contribution ($e^{-m^{2}s+ (\zeta-\frac{1}{6})R}$) for the R-summed Schwinger-Dewitt coefficient \cite{Parker6, Jack, Ferreiro})  and the fluctuation contribution \cite{BARVINSKY1990}. This enables us to write the imaginary part of the effective action in weak field limits as ({\it vide.} {\bf Appendix A})

\ben
Im(\Gamma)=&&\frac{1}{32\pi^2}\int d^4x \sqrt{-g}\Big[\frac{\pi}{2}(M^2)^2\Theta(-M^2)+\nonumber\\ && \pi \Theta(-M^2)\Big(\frac{1}{6}\big(\zeta-\frac{1}{5}\big)\Box R+\nonumber\\ && \frac{1}{180}\big(R_{\mu\nu\rho\sigma}R^{\mu\nu\rho\sigma}-R_{\mu\nu}R^{\mu\nu}\big)\Big)]\nonumber \\
\label{12}
\een
It is analogous to the phenomenon of alpha particle tunneling via a Coulomb barrier. Here, $R$ represents the Ricci scalar, $R_{\mu\nu}$ represents the Ricci tensor, and the quantity $R_{\mu\nu\rho\sigma}R^{\mu\nu\rho\sigma}$ is
the Kretschmann scalar. The Kretschmann scalar provides insight into the strength of gravitational forces and the curvature of spacetime. 

It is important to highlight a significant difference between our approach and the method employed in Ref. \cite{Wondark}. In their formulation, particle production was studied using a standard expansion of the heat kernel where each term in the series depends explicitly on the curvature invariants through the Schwinger-DeWitt coefficients $a_n$ ({\it viz.} \cite{Wondark}). In contrast, our method incorporates an interesting enhancement by using an $R$-summed Schwinger-DeWitt coefficient \cite{Parker6, Jack, Ferreiro}) where the contributions proportional to the Ricci scalar $R$ are resummed directly into an exponential factor within the heat kernel, rather than being treated perturbatively in the coefficients $\Tilde{a}_n $ ({\it viz.} {\bf Appendix A}). 

This summation technique ($R$-summed) treats the Ricci scalar curvature in a completely non-perturbative fashion while treating the other curvature term perturbatively, resulting in more intricate phenomena than the traditional Schwinger-Dewitt method. As a result, our approach naturally accommodates the creation of both massless and massive particles. For massless particles, production is driven primarily by terms involving higher-order curvature invariants like the Kretschmann scalar $R_{\mu\nu\rho\sigma}R^{\mu\nu\rho\sigma}$ which remain non-zero even in Ricci-flat spacetimes. In contrast, for massive particles, the incorporation of the resummed Ricci scalar (R-summed) in the exponential modifies the effective action to avoid the strong exponential suppression typically associated with mass-dependent creation rates in standard formulations. This results in a more comprehensive and physically realistic description of particle production processes in curved spacetime geometries, including black hole spacetimes and cosmic contexts.

By treating the Ricci scalar non-perturbatively, our method extends the applicability of the heat kernel formalism to scenarios where both the local curvature and particle mass play significant roles in determining the dynamics of quantum fields. This refinement not only enhances the accuracy of particle production estimates but also provides a unified framework for addressing massless and massive particle creation within a consistent geometric setting.

\subsection{Locally defined horizons and mass functions for (generalized) Vaidya Spacetime}

Now we are attempting to determine the particle creation rate for a generalized Vaidya spacetime (\ref{1}). To determine the rate at which particles are produced, we must do an integration across the volume where particles are marginally trapped. Due to the lack of a precise description of the event horizon in a dynamical spacetime, it becomes necessary to introduce the idea of a locally defined horizon for dynamical spacetime. Local horizons are often characterized by trapped surfaces, which are compact and orientable two-dimensional surfaces submerged in a four-dimensional environment. These surfaces have two distinct directions that are perpendicular to it, representing the paths of incoming and outgoing null rays. We need to explore the congruences of ingoing and outgoing null geodesics with tangent fields, $l^a$ and $n^a$, respectively, and how they propagate under strong gravity. We now define some fundamental concepts of surface linked to closed two-surfaces \cite{Booth, Ben-Dov, Penrose, Faraoni}:

\begin{enumerate}
\item[1.] A normal surface has $\Theta_l>0$ and $\Theta_n<0$ (a two-sphere in Minkowski space fulfills this feature).
\item[2.] A trapped surface is defined as $\Theta_l<0$ and $\Theta_n<0$. The outgoing, as well as the ingoing, future-directed null rays converge here rather than diverge, and outward-propagating light is pulled back by strong gravity.
\item[3.] A marginally outer trapped surface (MOTS) is defined as $\Theta_l=0$ (where $\Theta_l$ is the surface's outgoing null normal) with $\Theta_n<0$.
\item[4.] An untrapped surface is one with $\Theta_l\Theta_n<0$.
\item[5.] An anti-trapped surface corresponds to $\Theta_l>0$ and $\Theta_n>0$ (both outgoing and ingoing future-directed null rays are diverging).
\item[6.] A marginally outer trapped tube (MOTT) is a three-dimensional surface, which can be foliated entirely by marginally outer trapped (two-dimensional) surfaces. 
\end{enumerate}

Sawayama \cite{Sawayama} modified the conventional definition of the dynamical horizon, as outlined in the previous section, based on the interpretation of MOTS. Sawayama has shown that by using the revised definition of the dynamical 
horizon and dynamical horizon equation, the mass of a black hole would ultimately diminish, resulting in the transformation of the spacetime into Minkowski spacetime. This transformation occurs regardless of the original size of the 
black hole mass. Sawayama \cite{Sawayama} demonstrated that in the context of dynamic spacetime, it is necessary to consider the position of the MOTS. For the generalized Vaidya metric, we get the location of the horizon 
as $r_{D}$ and $r_{D}^{+}$. It is essential to figure out the significance of the two limit surfaces associated with $r_{D}$ and $r_{D}^{+}$. The inner surface $r_{D}$ may be seen as the innermost trapped surface from which the dynamical 
horizon evolves. This surface signifies the beginning of the dynamical horizon's existence. This phenomenon may occur when a black hole, which was previously stationary, transitions into a dynamic state as a result of external factors, such 
as the influx of matter. The outer surface $r_{D}^{+}$ represents the final marginally trapped surface when the dynamical horizon no longer exists, perhaps transforming into a stationary event horizon. This phenomenon may occur when a black 
hole ceases to accumulate mass. Additionally, it is important to note that Majumder et al. \cite{gm3} use the dynamical horizon equations derived from Sawayama's \cite{Sawayama} modified description of the dynamical horizon to quantify the 
mass loss caused by Hawking radiation in a generalized Vaidya-type geometry, however, in a different context.

In this context, we consider the adiabatic vacuum, which corresponds to a background spacetime that evolves gradually without experiencing rapid fluctuations, especially within the trapping surfaces. The evolution of the dynamical horizon depends on the mass function. For example, a trapped cylindrical surface is formed when the mass function varies linearly with $v$. Our aim is to investigate the dynamics of particle production within this trapped region in the generalized Vaidya spacetime using the effective action principle. The formation of a trapped horizon in Vaidya spacetime is characterized by the behavior of null expansions and the evolution of the mass function. Focusing on particle production within the trapped horizon region is particularly insightful because the dynamic geometry significantly influences quantum fields. In quantum field theory on curved spacetime, particle creation may arise from the time-dependent metrics that alter the vacuum state. In Vaidya spacetime, where the mass function $m(v,r)$ evolves with time as well as space, the resulting non-stationary gravitational background, and strong curvature effects amplify particle creation within the trapped region. This makes the trapped horizon an essential domain that may helps to explore non-equilibrium quantum processes associated with black hole formation and evaporation. Investigating particle production inside the dynamical horizon may provide a localized perspective on the Hawking effect, which may enrich our understanding of black hole thermodynamics in non-static, evolving spacetimes.\\

Now we write the generalized Vaidya metric (\ref{1}) as
\ben
ds^2 = -F(v,r) dv^2 + 2dvdr + r^2 d\Omega^2 ,
\label{13}
\een
with $F(v,r)=1-\frac{2m(v,r)}{r}$. From the generalized Vaidya metric (\ref{13}), we can compute the following quantities: the quadratic curvature invariant ($R_{\mu\nu}R^{\mu\nu}$), the Ricci scalar ($R$) and the Kretschmann scalar ($R_{\mu\nu\rho\lambda}R^{\mu\nu\rho\lambda}$):

\ben
R_{\mu\nu}R^{\mu\nu}&&=\frac{1}{2r^4}\Big[4+4F^2+8r^2F_{r}^2+8F+8rFF_{r}\nonumber\\&&+r^4F_{rr}^2+4rF_{r}\Big(-2+r^2F_{rr}\Big)\Big],
\label{14}
\een

\ben
R^2=\frac{1}{r^4}\Big(-2+2F+4rF_{r}+r^2F_{rr}\Big),
\label{15}
\een
and
\ben
R_{\mu\nu\rho\lambda}R^{\mu\nu\rho\lambda}= \frac{4(F-1)^2}{r^4}+\frac{4F_{r}^2}{r^2}+F_{rr}^2,
\label{16}
\een
where $F_r=\frac{\partial F}{\partial r}$ and $F_{rr}=\frac{\partial^2 F}{\partial r^2}$. The function $F(v,r)$ varies depending on the specific choice of the mass function $m(v,r)$. 

To find the particle production rate, we take the generalized mass as \cite{Vertogradov, Vertogradov5},
\ben
m(v,r)=C(v)+D(v)r^{1-2\omega},\nonumber\\
\omega\in [-1,1],~\omega\neq \frac{1}{2}.
\label{17}
\een
Here, $\omega$ is the equation of state parameter defined by the relation $P=\omega\rho$, where $P$ represents the pressure and  $\rho$ denotes the energy density of the type-II matter field, while $C(v)$ and $D(v)$ are arbitrary functions of the Eddington time $v$.

For our investigation, we now select three specific mass functions \cite{Vertogradov, Vertogradov5, Mkenyeleye, Dadhich, gm2}, which are discussed below:

{\bf Case 1:} $C(v)=\mu v$ and $D(v)=0$ with $\mu>0$ which gives,
\ben
m(v,r)=m(v)=\mu v,
\label{18}
\een
This case represents a linearly increasing mass with advanced time $v$, corresponding to a system continuously accreting null radiation or energy. It models scenarios such as a black hole growing due to surrounding matter infall or radiation influx, making it relevant for studying accretion-driven particle creation. In this case, the dynamic increase in gravitational energy leads to vacuum instability by increasing the curvature which affects the propagation of virtual particles, polarizing the vacuum and causing it to behave as a medium with fluctuating particle-antiparticle pairs \cite{Birrell, Mukhanov}. Such a mechanism can trigger particle production, similar to cosmological expansion-induced pair creation \cite{Raine}. In other words, we can say that the mass function (\ref{18}) provides a foundational example of a Vaidya spacetime, demonstrating how null radiation interacts with and modifies the geometry of spacetime in the context of Einstein's field equations.

{\bf Case 2:} $C(v)=\mu v$ , $D(v)=\nu$ ($constant$) and $\omega=0$ (dust case) with $\mu>0, ~\nu>0$ which gives,
\ben
m(v,r)=\mu v +\nu r,
\label{19}
\een
This model introduces a linear dependence on both time $v$ and radial coordinate $r$. The term $\mu r $ could represent a spatially varying energy density within a dust-like matter distribution. This case may apply to gravitational collapse scenarios involving matter fields (dust) in addition to radiation \cite{Birrell, Wald, Mukhanov4}. The interplay between time-dependent mass growth and spatial distribution can enhance curvature effects, making it relevant for localized gravitational pair production studies. The additional spatial dependence can enhance particle creation within the dynamically evolving horizons. So, we may say that the form of the mass function (\ref{19}) may provide the interplay of radiative accretion and spatially distributed matter. It provides an adaptable framework for investigating dust-dominated cosmic areas, gravitational collapse, and the possible generation of quantum particles in dynamically evolving spacetime geometries.

{\bf Case 3:} $C(v)=\mu v$ , $D(v)=-\frac{\mu^2 v^2}{2}$ and $\omega=1$ (stiff fluid case) with $\mu>0$ which gives,
\ben
m(v,r)=\mu v-\frac{\mu^2 v^2}{2r}.
\label{20}
\een
This function represents a mass that grows initially but includes a nonlinear correction proportional to $\frac{v^2}{r}$, modeling a stiff fluid configuration ($\omega=1$) \cite{Birrell,blau,Wald,Mukhanov4}. This setup is suitable for studying a system emitting radiation in a self-interacting, stiff fluid-like configuration, relevant for highly relativistic or extreme-energy conditions near black holes or collapsing stars. The quadratic time dependence introduces a more complex curvature evolution. This structure could drive more intense particle creation due to rapidly changing spacetime geometry at smaller scales.

These cases comprehensively explore varying dynamics of radiating systems in Vaidya geometry, providing a rich foundation for understanding gravitational particle production.

\subsection{Evaluation of the rate of particle production}
In this subsection, we will look into the particle production rate for the generalized Vaidya spacetime with the three mass functions listed above.

{\bf Case 1:} In this case, the mass function depends solely on time ($v$), similar to the standard Vaidya spacetime, where $m(v,r)\equiv m(v)$. With the specific definition of mass function (\ref{18}), we have $R=0$ and $R_{vv}=\frac{2}{r^2}\frac{\partial m}{\partial v}$, yet $R^{\mu\nu}R_{\mu\nu}\equiv R^{vv}R_{vv}=0$. As a result, the Kretschmann scalar term, $R_{\mu\nu\rho\sigma}R^{\mu\nu\rho\sigma}=\frac{48m^{2}(v)}{r^6}$, is the only existing contribution to the imaginary part of the effective action (\ref{12}).

By using the revised concept of the dynamical horizon and the accompanying methodology proposed by Sawayama \cite{Sawayama}, we get the values for the two dynamical horizon radii as $$r_D=2\mu v,~~\text{and}~~r_D^+=2\mu v(1+e^{-\frac{1}{\mu}}).$$ Substituting all the values of the above parameters (scalars) into the imaginary part of the effective action (\ref{12}), we obtain the rate of particle production in the dynamical Vaidya spacetime as
\ben
\frac{dN}{dv}&&=2\frac{dIm(\Gamma)}{dv} \nonumber\\&& =\frac{2}{64\pi}\int_{r_{D}}^{r_{D}^{+}}\int_0^{\pi}\int_0^{2\pi}r^2sin(\theta)drd\phi d\theta \frac{48m^{2}(v)}{180r^6}\nonumber\\&&
=\frac{m^{2}(v)}{90}\Big(\frac{1}{(r_{D})^{3}}-\frac{1}{(r_{D}^{+})^{3}}\Big)\nonumber\\&&=\frac{1}{720\mu v}\Big(1-\frac{1}{(1+e^{-\frac{1}{\mu}})^3}\Big).
\label{21}
\een

{\bf Case 2:} In the second scenario, where the mass function is given by $m(v,r)=\mu v +\nu r$, the dynamical horizon radius can be expressed as $$r_D=\gamma v, ~~\text{and}~~ r_D^+=\gamma v\Big(1+\frac{1}{W_0(\gamma v e^{-\frac{2\mu}{\gamma^2}})}\Big),$$ where $\gamma=\frac{2\mu}{1-2\nu}$ and $W_0$ represents the principal value of the Wright Omega function \cite{corless1, corless2, corless3}. The characteristics of the Wright Omega function may be found in the Ref. \cite{gm3}. In this case, all the terms in the imaginary part of the effective action will contribute, and which are given by: $R_{\mu\nu}R^{\mu\nu}=\frac{8\nu^2}{r^4}$, $R=\frac{4\nu}{r^2}$, $\Box R=-8\frac{4\mu \nu v+(2\nu^2-\nu)r}{r^5}$  and $K=R_{\mu\nu\rho\lambda}R^{\mu\nu\rho\lambda}=\frac{16}{r^6}\Big(\nu^2r^2+2\nu\mu v r+3\mu^2 v^2\Big)$ with the parameter $z$ as $z=\gamma v e^{-\frac{2\mu}{\gamma^2}}$. In this case, $R$ as well as $\Box R$, are not zero, which corresponds to both massive and massless particle production ({\it viz.} {\bf Appendix A}). Hence we analyze two cases corresponding to conformal ($\zeta=\frac{1}{6}$) and non-conformal ($\zeta <\frac{1}{6}$) coupling. For conformal coupling only massless particles will be produced; however, for nonconformal coupling, both massive and massless particles will be produced within a certain range of mass denoted by $0 \le m^2 < \big(\frac{1}{6}-\zeta \big) R$ with $\zeta <\frac{1}{6}$.\\

\textbf{A:} For conformal coupling($\zeta=\frac{1}{6}$) the particle production rate is expressed as:

\ben
&&\frac{dN}{dv}=2\frac{dIm(\Gamma)}{dv} \nonumber\\&& =\frac{2}{64\pi}\int_{r_{D}}^{r_{D}^{+}}\int_0^{\pi}\int_0^{2\pi}r^2sin(\theta)drd\phi d\theta \nonumber\\&&\frac{1}{180}\Big(R_{\mu\nu\rho\sigma}R^{\mu\nu\rho\sigma}-R_{\mu\nu}R^{\mu\nu}-\Box R \Big)\nonumber \\&&=\frac{1}{180}\int_{r_{D}}^{r_{D}^{+}}\Big(\frac{(3\nu^2-\nu) r^2+8\frac{\nu \mu}{\gamma} v r+6\frac{\mu^2}{\gamma^2} v^2}{r^4}\Big)dr \nonumber \\ &&
=\frac{1}{180\gamma v}\Big[\frac{(3\nu^2-\nu)}{1+W_0(z)}+4\frac{\nu\mu}{\gamma} \frac{1+2W_0(z)}{(1+W_0(z))^2}\nonumber \\&& +2\frac{\mu^2}{\gamma^2}\frac{1+3W_0(z)^2+3W_0(z)}{(1+W_0(z))^3}\Big].
\label{22}
\een

Note there is an important restriction that we must obey in order to have finite particle production which comes from the positivity of $\gamma$ and $(3\nu^2-\nu)$. For $\gamma>0$ we must take $\nu < \frac{1}{2}$ and for $(3\nu^2-\nu) \ge 0$ we require $\nu \ge \frac{1}{3}$. Thus, the range of $\nu$ that satisfies both conditions simultaneously is $\frac{1}{3} \le \nu < \frac{1}{2}$. Here we take $\nu=\frac{9}{22}$. \\

\textbf{B:} For non-conformal coupling ($\zeta=\frac{1}{8}$) the particle production rate depends on the mass of the produced particle. Here we consider the general case of particle production with mass $m^2=\frac{R}{\alpha}$ with $\alpha >24 $ ({\it viz.} {\bf Appendix A} Eq.{\ref{A21}} and discussion thereafter) and is expressed as:
\ben
&&\frac{dN}{dv}=2\frac{dIm(\Gamma)}{dv} \nonumber\\&& =\frac{2}{32\pi}\int_{r_{D}}^{r_{D}^{+}}\int_0^{\pi}\int_0^{2\pi}r^2sin(\theta)drd\phi d\theta \Big[\Big(\frac{1}{\alpha}-\frac{1}{24}\Big)^2\frac{R^2}{2} \nonumber\\&&+\frac{1}{180}\Big(R_{\mu\nu\rho\sigma}R^{\mu\nu\rho\sigma}-R_{\mu\nu}R^{\mu\nu}-\frac{9}{4}\Box R \Big)\Big]\nonumber \\&&=\frac{1}{2}\Big(\frac{1}{\alpha}-\frac{1}{24}\Big)^2\int_{r_{D}}^{r_{D}^{+}}\frac{\nu^2}{r^2}dr \nonumber \\ &&+\frac{1}{360}\int_{r_{D}}^{r_{D}^{+}}\Big(\frac{(22\nu^2-9\nu) r^2+52 \nu \mu v r+24\mu^2 v^2}{r^4}\Big)dr \nonumber \\ &&
=\frac{\nu^2}{2 \gamma v(1+W_0(z))}\Big(\frac{1}{\alpha}-\frac{1}{24}\Big)^2 +\frac{1}{360\gamma v}\Big[\frac{(22\nu^2-9\nu)}{1+W_0(z)}\nonumber \\ &&+26\frac{\nu\mu}{\gamma} \frac{1+2W_0(z)}{(1+W_0(z))^2} +8\frac{\mu^2}{\gamma^2}\frac{1+3W_0(z)^2+3W_0(z)}{(1+W_0(z))^3}\Big].\nonumber \\
\label{23}
\een

Note that there is also an important restriction, so as to ensure finite particle production which comes from the positivity of $\gamma$ and $(22\nu^2-9\nu)$. For $\gamma>0$ we must take $\nu < \frac{1}{2}$ and for $(22\nu^2-9\nu) \ge 0$ we require $\nu \ge \frac{9}{22}$. Thus, the range of $\nu$ that satisfies both conditions simultaneously is $\frac{9}{22} \le \nu < \frac{1}{2}$. Here for simplicity, we take $\nu=\frac{9}{22}$.

These two scenarios namely conformal and non-conformal coupling yield distinct particle production rates for massive and massless particles. In the second case, particles with masses upto $\frac{R}{24}$ can be produced, corresponding to the range  $0 \le m^2 \le \frac{R}{24}$. \\

{\bf Case 3:} The third scenario, characterized by the mass function $m(v,r)=\mu v-\frac{\mu^2 v^2}{2r}$, corresponds to the mass function of a charged-Vaidya type spacetime \cite{Mkenyeleye, Dadhich, gm2}.  The dynamical horizon radius may be expressed as follows: $r_D=\mu v$ and $r_D^{+}=\mu v\Big(1+\frac{B(v)}{W_0(B(v) e^{B^{2}(v)-2B(v)})}\Big)$. The expression $B(v)=\sqrt{\frac{1+4\mu+2\mu ln(\mu v)}{\mu}}$ defines the function $B(v)$, where $\mu$ is a constant. The function $W_0(z)$ represents the principal value of the Wright Omega function. It is important to ensure that the mass function remains positive within the dynamical horizon region (i.e. between $r_D$ and $r_D^+$). To examine the conditions under which the mass function could become negative, we analyze the inequality: $\mu v-\frac{\mu^2 v^2}{2r}<0$ to get $r<\frac{\mu v}{2}$. Since the radii of the dynamical horizons ($r_D$ and $r_D^+$) are always greater than the above value, therefore the positivity of the mass function is always ensured. Thus, we can say that the mass function will remain positive throughout the dynamical horizon, preserving the physical consistency of the horizon description. In this case, contribution comes from $R_{\mu\nu}R^{\mu\nu}$ and $R_{\mu\nu\rho\lambda}R^{\mu\nu\rho\lambda}$ to the imaginary part of the effective action (\ref{12}) and are provided by, $R_{\mu\nu}R^{\mu\nu}=\frac{4\mu^4 v^4}{r^8}$, $R=0$ and $R_{\mu\nu\rho\lambda}R^{\mu\nu\rho\lambda}=\frac{8}{r^8}\Big(7\mu^4v^4-2\mu^3rv^3+6\mu^2r^2v^2\Big)$. So the particle production rate for this scenario is,
\ben
\frac{dN}{dv}&&=2\frac{dIm(\Gamma)}{dv} \nonumber\\&& =\frac{2}{64\pi}\int_{r_{D}}^{r_{D}^{+}}\int_0^{\pi}\int_0^{2\pi}r^2 sin(\theta)drd\phi d\theta \nonumber\\&& \frac{1}{180}\Big(R_{\mu\nu\rho\sigma}R^{\mu\nu\rho\sigma}-R_{\mu\nu}R^{\mu\nu}\Big)\nonumber \\&& 
=\frac{1}{360}\int_{r_D}^{r_D^+}\Big(\frac{13\mu^4v^4-8\mu^3rv^3+12\mu^2r^2v^2}{r^6}\Big)dr \nonumber \\&&
=\frac{1}{360\mu v}\Big[\frac{13\mu^4}{5}\Big(1-\frac{1}{(1+\frac{B(v)}{W_0(B(v) e^{B^{2}(v)-2B(v)})})^5}\Big)\nonumber\\ &&
-2\mu^3\Big(1-\frac{1}{(1+\frac{B(v)}{W_0(B(v) e^{B^{2}(v)-2B(v)})})^4}\Big)\nonumber\\&& +4\mu^2\Big(1-\frac{1}{(1+\frac{B(v)}{W_0(B(v) e^{B^{2}(v)-2B(v)})})^3}\Big)\Big]
\label{23}
\een

The three cases discussed exhibit distinct characteristics. In the first and third scenarios, the coupling constant ($\zeta$) is irrelevant, resulting in the production of solely massless particles irrespective of its value, since the Ricci scalar is zero in these instances. Conversely, the second scenario permits the generation of both massive and massless particles, depending on the specific value of the coupling constant, as indicated above.

\subsection{Graphical explanation}
In this subsection, we discuss the particle production rate for the above three cases between $r_{D}$ and $r_{D}^{+}$.

We have constructed graphs showing the particle production rate as a function of advanced time $v$ for all three scenarios, using specific values of the constants that satisfy the energy conditions outlined in Refs. \cite{Vertogradov5, Wang, Nielsen, blau}. The corresponding energy-momentum tensor and energy conditions are given below. The energy-momentum tensor is of the form $T_{\mu\nu}=T_{\mu\nu}^{(n)}+T_{\mu\nu}^{(m)}$, where $T_{\mu\nu}^{(n)}=\sigma l_{\mu}l_{\nu}$ and $T_{\mu\nu}^{(m)}=(\rho+P)(l_{\mu}n_{\nu}+l_{\nu}n_{\mu})$ with $\sigma=\frac{2\dot{m}(v,r)}{r^{2}}$, $\rho=\frac{2m^{'}(v,r)}{r^{2}}$ and $P=-\frac{m^{''}(v,r)}{r}$, where $l_{\mu}$ and $n_{\mu}$ are two null vectors, $l_{\mu}=\delta^{0}_{\mu}$, $n_{\mu}=\frac{1}{2}\Big(1-\frac{2m(v,r)}{r}\Big)\delta^{0}_{\mu}-\delta^{1}_{\mu}$, $l_{\mu}l^{\mu}=0=n_{\mu}n^{\mu}$, $l_{\mu}n^{\mu}=-1$. The corresponding energy conditions are (a) {\it the strong and weak energy conditions:} $\sigma\geq 0;~\rho\geq 0;~P\geq 0~~(\sigma\neq 0)$, (b) {\it the dominant energy condition:} $\sigma\geq 0;~\rho\geq P\geq 0~~(\sigma\neq 0)$.

\begin{figure}[h]
\centering
        \includegraphics[width=1.2\linewidth]{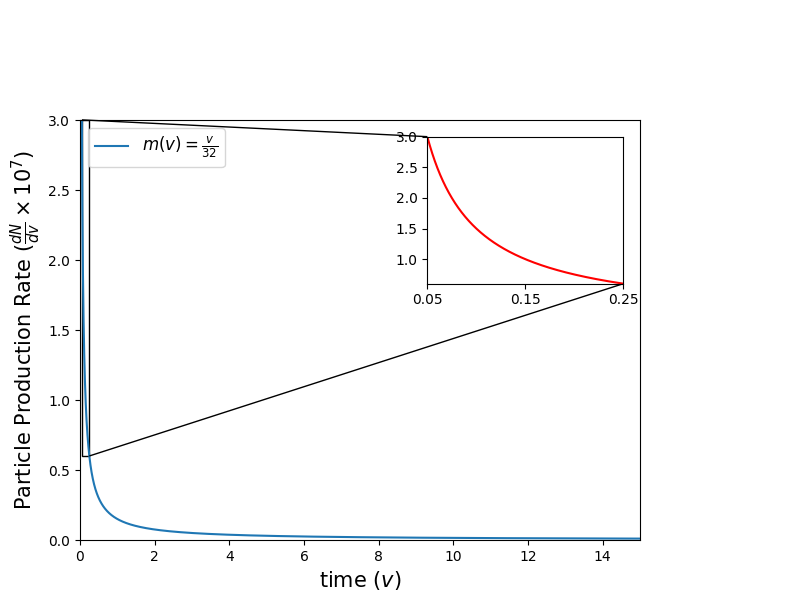}
        \caption{\textit{Particle production rate for $m(v)=\mu v$ with $\mu=\frac{1}{32}$}}
        \label{Ia}
    \end{figure}

\begin{figure}[h]
\centering
        \includegraphics[width=1.2\linewidth]{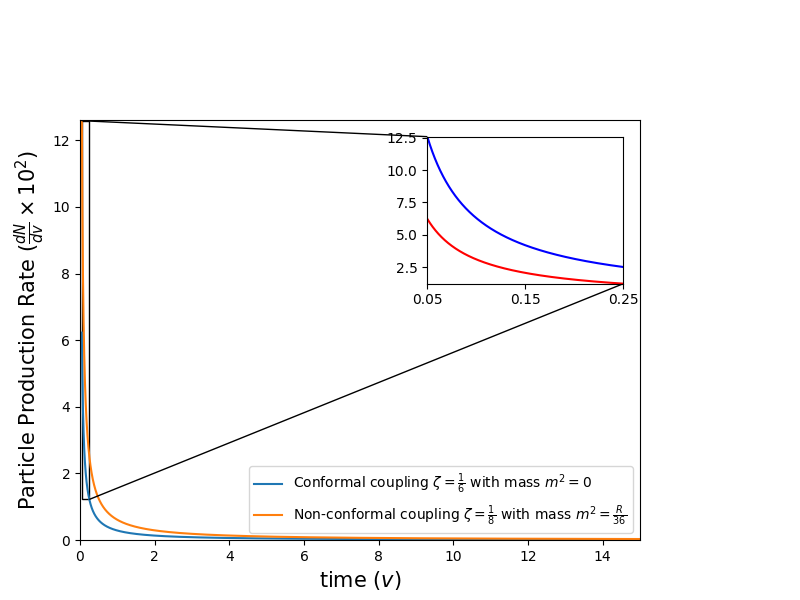}
        \caption{\textit{Particle production rate for $m(v,r)=\mu v+\nu r$ with $\mu=\frac{1}{24}$ and $\nu=\frac{9}{22}$}}
        \label{Ib}
    \end{figure}

\begin{figure}[h]
\centering
        \includegraphics[width=1.1\linewidth]{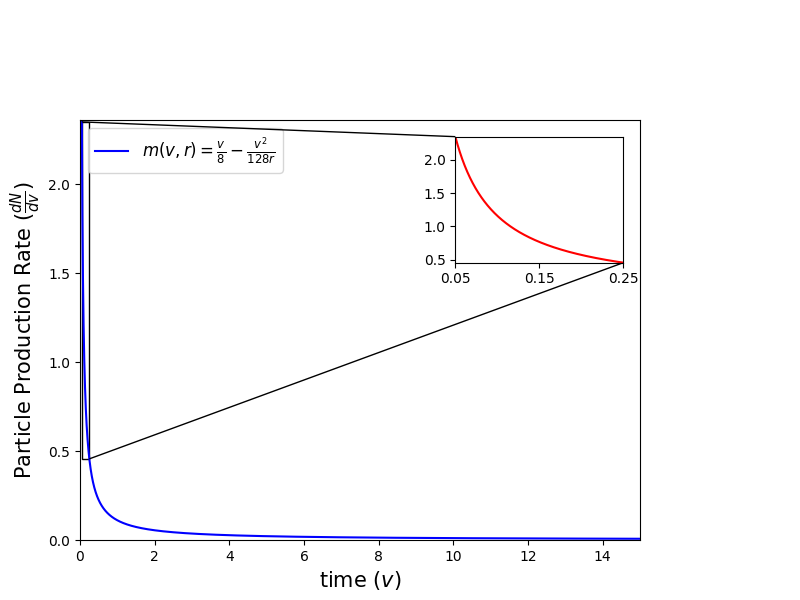}
        \caption{\textit{Particle production rate for $m(v,r)=\mu v-\frac{\mu^2 v^2}{2r}$ with $\mu=\frac{1}{8}$}}
        \label{Ic}
\end{figure}

The three plots (Figs. (\ref{Ia}), (\ref{Ib}) and (\ref{Ic}) ) illustrate that the particle creation rate decreases with time ($v$), finally approaching zero asymptotically in each case. This graphical representation shows only the finite portion of the graph. The three graphs depict varying particle production rates, each differing in order of magnitude. We can interpret it both mathematically and physically. Mathematically, the particle production rate can be understood as the difference in the number of particles produced between two surfaces, defined by the radii $r_D$ and $r_D^+$. Given that curvature quantities such as the Ricci scalar, Kretschmann scalar, and $R_{\mu\nu}R^{\mu\nu}$ exhibit scaling behavior with inverse powers of the radius, it follows that the rate of particle production increases as the separation between the two dynamical surfaces increases. This relationship subsequently leads to a significant order of magnitude difference in the production rate. Physically, these differences most likely stem from varying physical parameters or conditions influencing the rate of particle generation, such as the mass function, pressure, or energy density of the system.  The graphs collectively depict the hierarchy of magnitudes in particle production, highlighting the specific conditions that maximize the generation of particles.

The Fig. (\ref{Ia}) represents the lowest level of particle generation, signifying less interaction between the quantum field and the background geometry. This may occur because, in this instance, the mass function ($m(v,r)$) is just a function of $v$ and is independent of the spatial variable $r$.

The inclusion of radial dependence in the second scenario, which is the dust example (Fig. \ref{Ib}) with a generalized mass function $m(v,r)$ (\ref{19}), may increase the interaction between the quantum fields and the background spacetime geometry, which in turn leads to an increase in the rate of particle creation. There is a possibility that this dust situation, which is essentially pressure-less, would not significantly affect the geometry; yet, the radial dependency will still increase the production rate. Within this case, we analyze two subcases corresponding to conformal and non-conformal coupling. The graph illustrates that particle production in a dynamic background is strongly influenced by the type of coupling and particle mass. Conformal coupling leads to the production of massless particles, whereas non-conformal coupling allows for the generation of massive particles. The temporal evolution reveals that particle production is most prominent at early times and gradually decreases as time progresses.

The highest particle production rate occurs in the final scenario (Fig. \ref{Ic}), associated with the mass function $m(v,r)$ (\ref{20}) for a stiff fluid. This enhanced rate arises due to the fluid's pressure, which, along with the energy density, contributes significantly to particle generation. As pressure influences the spacetime geometry, it enhances the energy exchange within the system, ultimately resulting in the greatest particle production. Thus, it may be interpreted that the fluid's dynamic interplay of pressure and energy density is key to maximizing the particle creation rate.

Therefore, we can conclude that, based on the three figures shown above, if the mass function is dependent on both $v$ and $r$, especially when there is spatial dependency and considerable pressure (as in the case of stiff fluid), then the curvature effects are larger as well as the particle production rate is higher.

We are going to look at the physical implications of the particle generation rate resulting from the effects of quantum fields, as shown by the figures mentioned above. The graphs indicate an initial high rate of particle generation followed by a gradual decrease, suggesting that the early phase may represent a period of significant energy transfer or interaction between the quantum field and the evolving spacetime geometry. This may pertain to accretion processes in which the first infall of matter induces substantial spacetime fluctuations via quantum fields. As time progresses, the spacetime may approach a more stable state, and the rate of change in the geometry decreases. In our scenario, the Vaidya spacetime may be evolving into a quasi-static or less dynamic state, which could result in a decrease in particle production. 

An alternative explanation could be framed in terms of diminishing available energy, as the particle production itself may induce a backreaction on the spacetime, potentially slowing down the production rate over time. This occurs as the quantum field interacts with the background, effectively lowering the energy available for sustained particle creation. However, we do not address the backreaction effect in this work. Moreover, as the changes in the geometry slow down over time, the quantum field may evolve more adiabatically, resulting in a decrease in particle production as the system approaches equilibrium.

Alternatively, from Figs. (\ref{Ia}), (\ref{Ib}) and (\ref{Ic}), it follows that the discontinuous change in the mass function at $v\rightarrow 0$ leads to a sudden change in the spacetime background. This sudden change introduces a highly nonadiabatic evolution of the quantum field, which ultimately results in a significant surge in the particle production rate at the onset. However, as the geometry stabilizes, this rate decreases, indicating the progression of the system towards equilibrium or a lower energy state. Moreover, to ensure a finite particle production rate, we omit the discontinuity point $v=0$ explicitly. The underlying reason is that, at $v=0$ the mass function itself vanishes for the first and third cases, although in the second case, a nontrivial dependence on $r$ will remain. However, from our very definition of $F(v,r)=1-\frac{2m(v,r)}{r}$ it follows that for the second case, $F$ is independent of any dynamical variable ($r$ and $v$) at $v=0$ which indicated the absence of any interaction, resulting in zero particle production. Thus, the parameter $v$ serves as a key indicator of the system’s dynamical behavior and the particle production mechanisms rely on non-zero dynamical parameters ($v \ne 0$) to generate mass or facilitate interactions. By omitting, $v=0$ we avoid scenarios where the mass function ceases to have physical relevance and ensure consistency with non-zero dynamical processes. It is important to note that all three mass functions discussed earlier satisfy the energy conditions outlined in \cite{Vertogradov}. There are several works that show that when a quantum field interacts with curved spacetime, the dynamic nature of the mass function can cause particles to form without breaking any of the known energy conditions \cite{Parker1, Ford, Birrell, Hiscock}.

Our findings align closely with Hawking’s quantum tunneling mechanism, revealing a shared foundational process where virtual particle pairs transition into real particles through the influence of gravitational energy. In Hawking's framework \cite{Hawking1}, this phenomenon occurs at the event horizon of a black hole, where the intense curvature of spacetime enables one particle to escape to infinity while its counterpart falls into the black hole, resulting in a net radiation of energy as Hawking radiation.

In contrast, our model demonstrates a complementary mechanism of particle production driven by the interaction of quantum fields with a time-dependent curvature rather than a fixed horizon structure. The dynamic evolution of spacetime curvature acts as a source of energy that destabilizes vacuum fluctuations, allowing virtual particles to gain real energy and become observable. This curvature-induced production mechanism underscores a broader generalization of quantum field processes in non-static geometries, that highlights particle creation is a fundamental consequence of gravitational field dynamics, not confined solely to horizons but also may present in evolving cosmological and radiating black hole like scenarios.\\

\subsection{Thermodynamics of the dynamic spacetime}
This section explores the thermodynamics of the generalized Vaidya spacetime, with a particular emphasis on examining surface gravity in regions near marginally outer trapped surfaces through surface gravity. Specifically, to understand the connection between 
gravity and thermodynamics in nonstatic systems, we focus on the surface gravity associated with dynamical horizons. 

In black hole thermodynamics, the surface gravity serves a role similar to that of temperature. However, in a fully dynamical scenario, surface gravity is not directly equivalent to the temperature of any thermal spectrum. Still, it is expected 
to play a significant role in the emission of Hawking-like radiation, even in non-equilibrium processes. Surface gravity is traditionally defined on a Killing horizon, which works well for stationary cases, but this approach fails in dynamical 
situations where no Killing horizon exists. Therefore, to determine the surface gravity for the dynamical Vaidya metric, we must use the Kodama vector field. This field, in the context of a spherically symmetric dynamical spacetime, provides a 
way to define a conserved current and energy when a global timelike Killing vector is absent \cite{Kodama, Nielsen, Faraoni}. So, we can say that the Kodama vector is an extension of the concept of the Killing vector field applicable to 
spacetimes that lack a timelike Killing vector, and it has been used in the thermodynamics of dynamically evolving horizons. To define the Kodama vector, we write the Vaidya metric (\ref{1}) in the form,
\ben
ds^2=h_{ab}(x_c)dx^a dx^b+\Tilde{r}^2(d\theta^2+sin^2 \theta d\phi^2)
\label{24}
\een
and the Kodama vector $K^{a}$ is defined as,
\ben
-\epsilon^{ab}\nabla_{b}\Tilde{r}=K^{a}
\label{25}
\een
with $K^{\theta}=K^{\phi}=0$, where $(a,b)\equiv (v,\Tilde{r})$, $\Tilde{r}$ is the areal radius, $\epsilon^{ab}$ is the two dimensional levi-Civita tensor in the $(v,\Tilde{r})$ space and $\nabla_b$ is the covariant derivative with respect to the metric $h_{ab}$ defined as 
$h_{vv}=-(1-\frac{2m(v,r)}{r})$, $h_{vr}=h_{rv}=1$ and $h_{rr}=0$. The Kodama vector satisfies the relation $K^{a}\nabla_{a}\Tilde{r}=-\epsilon^{ab}\nabla_{a}\Tilde{r}\nabla_{b}\Tilde{r}=0$. For the Vaidya metric, the Kodama vector becomes $K^a=(-1,0,0,0)$ \cite{Faraoni, Kodama}. The surface gravity $\kappa$ 
at the horizon can be defined in terms of Kodama vector as \cite{Faraoni},
\ben
K^b\nabla_b K^a=\kappa K^a.
\label{26}
\een
The definition of surface gravity ($\kappa$) for any dynamical spacetime can be found as~\cite{Kodama,Criscienzo,Faraoni,Sherif}

\ben
\kappa=\frac{1}{2\sqrt{-h}}\partial_{a}(\sqrt{-h}h^{ab}\partial_{b}\Tilde{r}),
\label{27}
\een
where the associated temperature named after Hayward-Kodama temperature at the apparent horizon is $T=\frac{|\kappa|}{2\pi}$. 

Our present study involves two dynamical horizons, namely $r_{D}$ and $r_{D}^{+}$. Thus, we have two surface gravities, referred to as inner ($\kappa_{1}$) and outer ($\kappa_{2}$), corresponding to two horizons that reflect the MOTS, which are provided by \cite{Faraoni}

\ben
\kappa_1=-\frac{1}{2}\frac{\partial}{\partial \Tilde{r}}g_{vv}|_{\Tilde{r}=r_{D}}&&\nonumber\\
\kappa_2=-\frac{1}{2}\frac{\partial}{\partial \Tilde{r}}g_{vv}|_{\Tilde{r}=r_{D}^+}
\label{28}
\een

Note that the temperature of a radiating star determines its energy emission spectrum, implying that higher (or lower) surface gravity corresponds to higher (or lower) temperatures, resulting in greater (or lesser) emission of radiation. We now aim to determine surface gravity ($\kappa$) for the three different cases under investigation.\\

{\bf Case I:} $m(v,r)=\mu v$
\ben
\kappa_1=\frac{1}{4\mu v};
\kappa_2=\frac{1}{4\mu v(1+e^{-\frac{1}{\mu}})^2}
\label{29}
\een

{\bf Case II:} $m(v,r)=\mu v +\nu r$
\ben
&&\kappa_1=\frac{1-2\nu}{2\gamma v};  \nonumber \\ &&
\kappa_2=\frac{(1-2\nu)W_0^{2}(z)}{2\gamma v(1+W_0^{2}(z))}
\label{30}
\een

\begin{figure}[H]
\centering
    \includegraphics[width=1\linewidth]{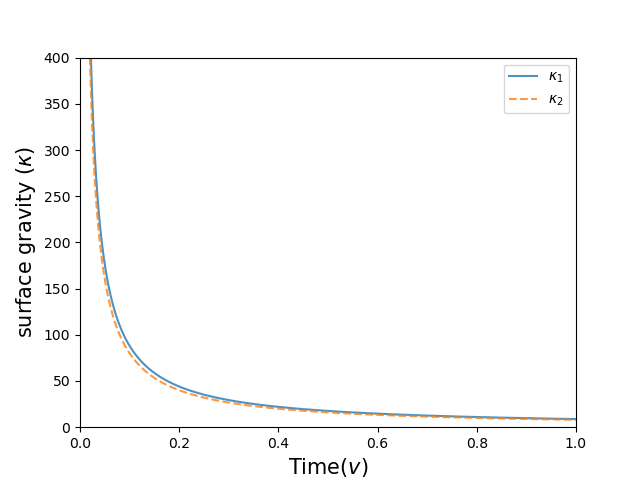}
        \caption{\textit{Surface gravity ($\kappa$) vs. Time (v) graph for $m(v)=\mu v$ with $\mu=\frac{1}{32}$}}
        \label{IIa}
    \end{figure}

\begin{figure}[h]
\centering
    \includegraphics[width=1\linewidth]{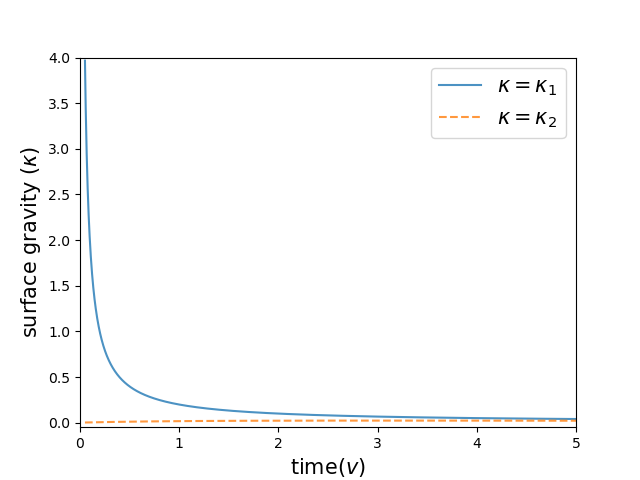}
        \caption{\textit{Surface gravity ($\kappa$) vs. Time (v) graph for $m(v,r)=\mu v+\nu r$ with $\mu=\frac{1}{24}$ and $\nu=\frac{9}{22}$}}
        \label{IIb}
    \end{figure}

\begin{figure}[h]
\centering
    \includegraphics[width=1\linewidth]{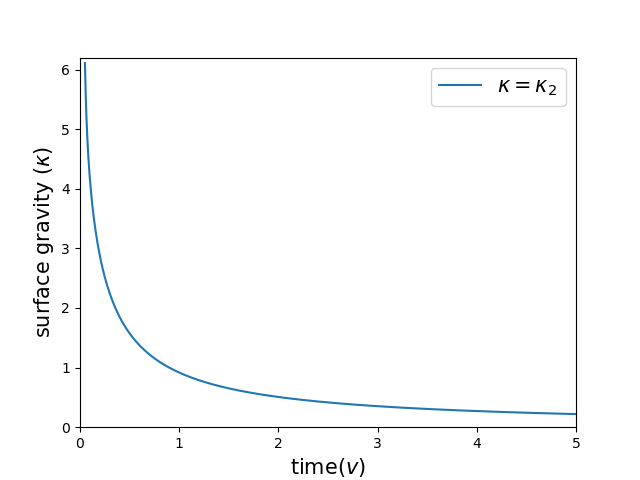}
        \caption{\textit{Surface gravity ($\kappa$) vs. Time (v) graph for $m(v,r)=\mu v-\frac{\mu^2 v^2}{2r}$ with $\mu=\frac{1}{8}$}}
        \label{IIc}
\end{figure}

{\bf Case III:} $m(v,r)=\mu v-\frac{\mu^2 v^2}{2r}$
\ben
&&\kappa_1=0;       \nonumber \\ &&
\kappa_2=\frac{1}{\mu v}\frac{\frac{B(v)}{W_0(B(v)e^{B^{2}(v)-2B(v)})}}{(1+\frac{B(v)}{W_0(B(v)e^{B^{2}(v)-2B(v)})})^3}
\label{31}
\een
The parameters $\gamma$, $z$, and $B(v)$ are defined as previously mentioned. We draw the graph of surface gravity($\kappa$) {\it vs.} time ($v$) for all three cases.  Note that in the first two instances ({Figs. \ref{IIa} and \ref{IIb}}), the surface gravity of the inner horizon exceeds that of the outer horizon owing to the stronger gravitational pull towards the core, where spacetime curvature is more intense. This results in a steeper gradient in the gravitational field at the inner horizon, causing its surface gravity to be greater. However, in the third case ({Fig.\ref{IIc}}) , the inner horizon exhibits zero surface gravity, as seen in extremal Reissner-Nordström or extremal Kerr black holes. Here, the inner horizon becomes degenerate, reflecting a precise balance in gravitational forces that neutralize the surface gravity. Meanwhile, the outer horizon retains a positive surface gravity, signifying that it remains non-degenerate and is still capable of trapping particles and radiation effectively.

From the plots (Figs. \ref{IIa}, \ref{IIb} and \ref{IIc} ), we can see that surface gravity decreases as time progresses, and eventually the system reaches equilibrium as the two surface gravity approaches equality. This reduction in surface gravity indicates that the gravitational pull at the horizon weakens over time. Physically, this suggests that the horizon's ability to retain particles or radiation diminishes, effectively slowing down the particle production rate, which supports our results. Since surface gravity is related to temperature, thermodynamically, the decrease in temperature implies that the horizon is cooling over time. This indicates that the system is transitioning to a lower energy state, most likely as a result of the accretion of mass into the radiating star, which spreads gravitational energy over a larger horizon area and lowers the temperature. This behavior can be matched as a radiating star gradually reaching equilibrium, with profound implications for the interplay between horizon thermodynamics, quantum fields, and the background geometry.

It is important to clarify that our focus is not on the final state after collapse, but rather on the collapse process itself. If we consider the gravitational collapse under the framework of the generalized Vaidya geometry, we may have two following scenarios:

If the gravitational collapse leads to the formation of a black hole, which may characterized by the emergence of a dynamical horizon that eventually evolves into a well-defined event horizon as the mass function stabilizes. The presence of a trapped surface signifies the development of a horizon, while the gradual decrease in surface gravity ($\kappa$) indicates a transition toward equilibrium, reinforcing the standard picture of black hole formation.

While a naked singularity can theoretically form under special conditions—such as a temporary alignment of the Cauchy horizon with the evolving event horizon—these scenarios are typically unstable. Quantum effects, back reaction, and small perturbations usually lead to the eventual hiding of the singularity within a horizon.

But, if a naked singularity does form, particle creation is still possible due to the strong gravitational field. Unlike black hole radiation, which is governed by the event horizon, a naked singularity could allow high-energy quantum emissions to escape, potentially leading to intense radiation (i.e only massless particle production) \cite{Parker10, Miyamoto}. However, such configurations are generally transient and do not represent the dominant outcome in our analysis.

Therefore, the final state of collapse— whether a black hole or a naked singularity— is of secondary importance in the context of our study, as both scenarios allow for particle production, albeit through different mechanisms. Our primary focus remains on the evolution of the dynamical horizon and its impact on particle production within the trapping region, rather than on the ultimate fate of the collapsing system.

\section{Conclusion}

This study analyzed the rate of particle formation and the thermal spectrum of the generalized Vaidya geometry by evaluating three different mass functions: $m(v)=\mu v$, $m(v,r)=\mu v +\nu r$, and $m(v,r)=\mu v-\frac{\mu^2 v^2}{2r}$. By analyzing the effective action, we successfully identified a clear relationship between particle creation rates and the dynamic evolution of spacetime within the MOTS, which is enclosed by two distinct dynamical horizons. In every case, we found that the particle creation rate starts off extremely high but decreases quickly over time, eventually approaching zero asymptotically. This pattern indicates the dynamic nature of particle production in these space-time geometries, where early-time conditions dominate the creation process, followed by a sharp decline as the system progresses.

The thermodynamic features of these instances further illustrate a complicated interaction between matter and geometry.  The findings indicate that the configuration of the mass function in Vaidya spacetime significantly affects particle generation and the corresponding thermodynamic properties. The graphical investigation of surface gravity over time emphasized the variations in surface gravity behavior across various scenarios, demonstrating the distinct thermodynamic evolutions based on the mass function.

Our findings enhance the understanding of the creation of particles in dynamic spacetime geometries, providing insights relevant to quantum field theory in curved spacetime and the thermodynamics of any dynamic spacetime. Subsequent investigations might look into the implications of intricate mass functions and dynamic spacetime geometries, along with the backreaction of particle generation on evolving spacetime. However, in this paper, we haven't taken into account the backreaction effects of the fields, as it lies beyond the scope of this study.

The study reveals that the Ricci scalar $R$ disappears in the first and third cases of the $R$-summed Schwinger-DeWitt method \cite{Parker6, Jack, Ferreiro}, leading to massless particle production only. The production of massive particles is exponentially suppressed in these geometries due to the absence of a non-zero Ricci scalar to counterbalance the mass-energy threshold. In contrast, the second case—where the metric is not Ricci-flat—enables the production of massive as well as massless particles, as the non-zero Ricci scalar contributes directly to the effective action, lowering the energy barrier for particle creation.
In Ref. {\cite{Wondark}} gravitational pair production was investigated for the Schwarzschild spacetime and as they show by using the Schwinger-Dewitt method they can only produce massless particles. 

The above circumstances are widely different from the case of Hawking radiation from black holes, where both massless and massive particles are emitted regardless of the specific spacetime curvature as long as the mass of an emitted particle is sufficiently smaller than the inverse Schwarzschild radius. Hawking radiation depends on the global properties of the event horizon and quantum field tunneling effects rather than local curvature invariants alone. As a result, Hawking's methodology is not subject to the same constraints as our perturbative analysis of dynamical spacetime, where dynamical horizons are present.

While our findings indicate important insights into the dynamics of particle creation in curved spacetime, there remain some limitations in our perturbative treatment of the effective action that we should address. The primary challenge with our method lies in its reliance on a perturbative expansion of the heat kernel, where curvature-dependent terms (except Ricci scalar) are treated separately and not resummed into a form that could capture non-local effects or the production of massive particles even in Ricci-flat backgrounds. Thus, developing a comprehensive framework within the path integral formalism that accurately captures particle creation processes analogous to Hawking radiation, while ensuring consistency with both local and global spacetime properties, remains an open challenge.

By addressing these limitations, we can deepen our understanding of massive particle creation mechanisms in the early universe and near black holes, ultimately bridging the gap between different approaches to quantum field behavior in curved spacetime. 

\vspace{0.3in}
{\bf  Acknowledgements:}

The authors would like to thank the referees for illuminating suggestions for improving the paper. G.M. would like to extend thanks to all the undergraduate, postgraduate, and doctoral students who significantly enriched him.

\appendix
\section{Effective Action Calculation}
The Action of the scalar field is given as,
\ben
S[\phi]=&&\int d^4x \sqrt{-g}[\frac{1}{2}g^{\mu \nu}\partial_{\mu}\phi\partial_{\nu}\phi- \nonumber\\&& \frac{1}{2}m^2\phi^2-\frac{1}{2}\zeta R\phi^2]
\label{A1}
\een
To find the effective action we need to integrate out the scalar field in the path integral formulation of quantum field theory,
\ben
e^{i\Gamma[g_{\mu\nu}]}=\int \mathcal{D}\phi e^{iS[\phi,g_{\mu\nu}]}
\label{A2}
\een
by analytic continuation we get euclidean effective action as,
\ben
e^{-\Gamma_E[g_{\mu\nu}]}=\int \mathcal{D}\phi_E e^{-S_E[\phi,g_{\mu\nu}]}
\label{A3}
\een

\ben
=\int \mathcal{D}\phi_E e^{-\int d^4x_E\sqrt{g_E}[\frac{1}{2}g_E^{\mu\nu}\partial_{\mu}\phi_E\partial_{\nu}\phi_E-\frac{1}{2}m^2\phi_E^2-\frac{1}{2}\zeta R\phi_E^2]}\nonumber
\een

after re calibrating we can write,
\ben
=\int \mathcal{D}\phi_E e^{-\int d^4x_E\frac{\sqrt{g_E}}{2}(\phi_E[-\Box_E+m^2+\zeta R]\phi_E)}
\label{A4}
\een
Thus we find the Euclidean one-loop effective action as,
\ben
\Gamma_E[g_{\mu\nu}]=\frac{1}{2}ln (\hspace{0.1cm} det[-\Box_E+m^2+\zeta R])\nonumber \\ 
\label{A5}
\een

Now here we use zeta function regularisation with complex parameter z and an arbitrary renormalization mass scale $\Tilde{\mu}$ to keep proper physical dimension \cite{Wondark}, where we take,
\ben
\Gamma_{E}[g_{\mu\nu}]=-\frac{{\Tilde{\mu}}^{2z}}{2}\int_0^{\infty}\frac{ds}{s^{1-z}}e^{-s(m^2+\zeta R-i\epsilon)}Tr \hat{K}(s) \nonumber \\ 
\label{A6}
\een

The Kernel K($x,x';s $) satisfies the equation,
\ben
(\partial_s -\Box_E) K(x,x';s)=0 
\label{A7}
\een
where $tr K(x,x';s)$ is the functional trace with the initial condition,
\ben
K(x,x';0)=\frac{\delta^4(x-x')}{\sqrt{g_E}}
\label{A8}
\een

The R-summed Schwinger-Dewitt expansion or the proper time expansion following Refs. (\cite{Wondark, Parker6}), is given as
\ben
<x|\hat{K}(s)|x>=e^{\frac{R s}{6}}\frac{\sqrt{g_E}}{(4\pi s)^2}\sum_{n=0}^{\infty}\Tilde{a}_n(x)s^{n} 
\label{A9}
\een
The R-summed Schwinger-DeWitt coefficients,$\Tilde{a}_n(x)$	are local scalar functions of the curvature tensor $R_{\mu\nu\lambda\kappa}$
that represents the coincidence limit of the heat kernel expansion. This expansion does not assume the curvature to be small and provides a formal method to express effective action in curved spacetime. In particular, we are interested in the proper time expansion up to the second order. These coefficients encapsulate the local geometric properties of spacetime and are essential in quantum field theory in curved backgrounds. Note that here we use the R-summed Schwinger-Dewitt coefficient. Following Parker and Toms \cite{Parker6}, we can write the Schwinger Dewitt coefficient up to second order as,
\ben
\Tilde{a}_0(x)=1
\label{A10}
\een
\ben
\Tilde{a}_1(x)=0 
\label{A11}
\een
\ben
\Tilde{a}_2(x)=&&\frac{1}{6}\Big(\zeta-\frac{1}{5}\Big)\Box R+\frac{1}{180}\Big(R_{\mu \nu \rho \sigma}R^{\mu \nu\rho \sigma}-R_{\mu \nu}R^{\mu \nu}\Big)  \nonumber \\ 
\label{A12}
\een

where $\Box$ denotes the Laplacian operator in 4 dimensions. Following the work of Refs. \cite{Parker6, Jack} we can rewrite the heat kernel trace as,
\ben
&& Tr \hat{K}(s)= K(x,x;s)\approx \nonumber \\ && e^{\frac{R s}{6}} \frac{1}{(4\pi s)^2}\int d^4x \sqrt{g_E} \Big[1+s^2\Big(\frac{1}{6}\big(\zeta-\frac{1}{5}\big)\Box R \nonumber \\ &&+\frac{1}{180}\big(R_{\mu \nu \rho \sigma}R^{\mu \nu\rho \sigma}-R_{\mu \nu}R^{\mu \nu}\big)\Big)\Big]+O(s^3) \nonumber \\
\label{A13}
\een

With the classical contribution to the heat kernel $e^{-sm^2}$ and the non-perturbative Ricci scalar contribution $e^{- \zeta R s}$ , we insert the above expression in the effective action (\ref{A6}) as follows:
\ben
\Gamma_E = && -\frac{\Tilde{\mu}^{2z}}{32\pi^2} \int d^4x \sqrt{g_E}\nonumber \\ && \int \frac{ds}{s^{3-z}} \sum_{n=0}^{\infty}\Tilde{a}_n(x)s^{n} e^{-s(M^2-i\epsilon)}\nonumber \\ 
\label{A14}
\een
where we write $M^2=m^2+(\zeta-\frac{1}{6})R$ and we introduced a regulator $i\epsilon$ in order to eventually take the limit $M\to 0$.
If we focus on the lowest order in the curvature tensors $R_{\mu\nu\rho\lambda}$, $R_{\mu\nu}$ and $R$, while ignoring terms involving their derivatives (which are associated with higher powers in the proper time expansion), we can approximate the effective action without including the more complex derivative contributions. This approach simplifies the expression by retaining only the leading curvature-dependent terms and discarding those tied to more subtle geometric variations of the spacetime manifold.

Integrating the above equation w.r.t $s$, we have
\ben
\int_0^{\infty} \frac{ds}{s^{1-z}}s^j e^{-s(M^2-i\epsilon)}= \Gamma(j+z)(M^2-i\epsilon)^{-j-z}\nonumber \\ 
\label{A15}
\een
where we use $\int_0^{\infty}e^{-st}t^{n-1} dt=\Gamma(n)s^{-n}$. Here $j=n-2$ and the effective action will have an imaginary part only if it $j$ takes a negative integer value or zero ($j \le 0 $). For $j= -2,-1,0,...$ corresponding to the first three terms of the expansion ($n=0,1,2$) of Eq. (\ref{A14}) and appropriate values of $z$, we get,
\ben
\Gamma_E=&&-\frac{\Tilde{\mu}^{2z}}{32\pi^2}\int d^4x\sqrt{g_E}\Big[\Gamma(-2+z)(M^2-i\epsilon)^{2-z} \nonumber \\ && +\Gamma(z)(M^2-i\epsilon)^{-z}\Big(\frac{1}{180}(R_{\mu\nu\rho\sigma}R^{\mu\nu\rho\sigma}\nonumber \\ &&-R_{\mu\nu}R^{\mu\nu}) +\frac{1}{6}\big(\zeta-\frac{1}{5}\big)\Box R\Big)\Big]+...
\label{A16}
\een
\subsection{Particle Production in Gravitational Fields}
In Euclidean signature, the scalar particle production rate is given by the imaginary part of the effective action,$-2Im(\Gamma_E)$. This imaginary part arises due to the branch cut in the logarithmic term (taken along the negative real axis) within the expression $(M^2-i\epsilon)^{-z}$. The branch cut signals instability in the field, which is responsible for particle creation, and the imaginary part represents the probability amplitude for these quantum tunneling events. Now, we have
\ben
\Tilde{\mu}^{2z}(M^2-i\epsilon)^{-z}=&&1-zln(\frac{M^2}{\Tilde{\mu}^2}-i\epsilon)+\mathcal{O}(z^2)\nonumber \\ 
\label{A17}
\een
where 
\ben
ln((\frac{M^2}{\Tilde{\mu}^2}-i\epsilon)&&=ln\Big|\frac{M^2}{\Tilde{\mu}^2}-i\epsilon\Big|+iArg\Big(\frac{M^2}{\Tilde{\mu}^2}-i\epsilon\Big)\nonumber \\ &&=ln\Big|\frac{M^2}{\Tilde{\mu}^2}-i\epsilon\Big|-i\pi\Theta(-M^2)
\label{A18}
\een

In terms of the heaviside step function $\Theta$ \cite{Wondark},
\ben
\Theta(x)= \begin{cases} 
      0 & x < 0 \\
      \frac{1}{2} & x = 0\\
      1 & x>0
      \end{cases}
\label{A19}
\een
The value $\Theta=\frac{1}{2}$ in the origin is justified by the fact that Arg$(-i\epsilon)=-\frac{\pi}{2}$ for $M^2=0$ and for $M^2<0$ there remain an imaginary part for the above expression.

Combining this with the Laurent series of the $\Gamma$- function
\ben
\Gamma(-2+z)=\frac{1}{2z}+\frac{3}{4}-\frac{\gamma_E}{2}+\mathcal{O}(z),\nonumber \\
\Gamma(-1+z)=-\frac{1}{z}-1+\gamma_E+\mathcal{O}(z),\nonumber\\
\Gamma(z)=\frac{1}{z}-\gamma_E+\mathcal{O}(z)
\label{A20}
\een
We only collect terms that is constant in z to find the imaginary part of the re-normalized effective action. After Wick rotating, the effective action to the Lorentzian spacetime we may write it as,
\ben
Im(\Gamma)=&&\frac{1}{32\pi^2}\int d^4x \sqrt{-g}\Big[\frac{\pi}{2}(M^2)^2\Theta(-M^2)+\nonumber\\ && \pi \Theta(-M^2)\Big(\frac{1}{6}\big(\zeta-\frac{1}{5}\big)\Box R+\nonumber\\ && \frac{1}{180}\big(R_{\mu\nu\rho\sigma}R^{\mu\nu\rho\sigma}-R_{\mu\nu}R^{\mu\nu}\big)\Big)]\nonumber \\ && +...
\label{A21}
\een
where dots stand for the higher curvature terms. We use this imaginary part of the effective action in our calculation of the rate of particle production for Vaidya geometry. The above result provides us with both massless (for conformal coupling $\zeta=\frac{1}{6}$) and massive particle production ($M^2<0$ or $m^2<\big(\frac{1}{6}-\zeta\big)R$). While for the first and third cases, it does not matter if the field is conformally coupled or not but for the second case, we study two cases separately corresponding to $\zeta=\frac{1}{6}$(conformal coupling) and $\zeta=\frac{1}{8}$( non-conformal coupling). This allows us to study both massless as well as massive particle production. Note that to ensure the positivity of the particle mass we always take $\zeta<\frac{1}{6}$. Therefore, we can write the imaginary part of the effective action for conformal($\zeta=\frac{1}{6}$) as well as non-conformal coupling($\zeta=\frac{1}{8}$) as:

\ben
Im(\Gamma)=
    &&\frac{1}{64\pi}\int d^4x \frac{\sqrt{-g}}{180}\Big[R_{\mu\nu\rho\sigma}R^{\mu\nu\rho\sigma}\nonumber \\ &&-R_{\mu\nu}R^{\mu\nu}-\Box R \Big] \nonumber
\een
for conformal coupling.
\ben
Im(\Gamma)=
    &&\frac{1}{32\pi}\int d^4x \frac{\sqrt{-g}}{180}\Big[R_{\mu\nu\rho\sigma}R^{\mu\nu\rho\sigma}\nonumber \\ &&-R_{\mu\nu}R^{\mu\nu}-\frac{9}{4}\Box R \Big]
\label{A22}
\een
for non-conformal coupling.

\end{document}